\documentstyle[prb,aps,12pt]{revtex}

\begin{document}
\title{Modelling of strain effects in manganite films}
\author{C. A. Perroni, V. Cataudella, G. De Filippis, G. Iadonisi, V. Marigliano, and F. Ventriglia}
\address{Coherentia-INFM  and Dipartimento di Scienze Fisiche, \\
Universit\`{a} degli Studi di Napoli ``Federico II'',\\
Complesso Universitario Monte Sant'Angelo,\\
Via Cintia, I-80126 Napoli, Italy}
\date{\today}
\maketitle

\begin{abstract}
Thickness dependence and strain effects in films of
$La_{1-x}A_xMnO_3$ perovskites are analyzed in the colossal
magnetoresistance regime. The calculations are based on a
generalization of a variational approach previously proposed for
the study of manganite bulk. It is found that a reduction in the
thickness of the film causes a decrease of critical temperature
and magnetization, and an increase of resistivity at low
temperatures. The strain is introduced through the modifications
of in-plane and out-of-plane electron hopping amplitudes due to
substrate-induced distortions of the film unit cell. The strain
effects on the transition temperature and transport properties are
in good agreement with experimental data only if the dependence of
the hopping matrix elements on the $Mn-O-Mn$ bond angle is
properly taken into account. Finally variations of the
electron-phonon coupling linked to the presence of strain turn out
important in influencing the balance of coexisting phases in the
film.
\end{abstract}

\newpage
The perovskite oxides $La_{1-x}A_xMnO_3$ ($A$ stands for a
divalent alkali element such as $Sr$ or $Ca$) have been studied
intensively since the discovery of "colossal" magnetoresistance
($CMR$) in thin films. \cite{helmolt} These dramatic changes in
electron and magnetic properties are found at temperatures around
the combined ferromagnetic-paramagnetic and metallic-insulating
($MI$) transitions. The ferromagnetic phase was usually explained
by introducing the double exchange mechanism, \cite{2} in which
hopping of an outer shell electron from a $Mn^{3+}$ to a $Mn^{4+}$
site is favored by a parallel alignment of the core spins. In
addition to the double-exchange term that promotes hopping of the
carriers, a strong interaction between electrons and lattice
distortions has been proposed to play a non negligible role in
these compounds \cite{3,millis} and confirmed by many experimental
measurements. \cite{5,6,7} Actually for the $Mn^{3+}$ site, with
three electrons in the energetically lower spin triplet state
$t_{2g}$ and the mobile electron in the higher doublet $e_g$, a
Jahn-Teller distortion of the oxygen octahedron can lead to
splitting of the doublet and the trapping of the charge carriers
in a polaronic state. A second important connection between
crystal structure and $MI$ transition lies in the dependence of
the $Mn-Mn$ electron transfer on the $Mn-O-Mn$ bond angle
$\theta$, that is on the orientation of the oxygen octahedra with
respect to the main crystal axes. This implies a strong effect of
either external pressure or mean $A$-site ionic radius on the
ferromagnetic critical temperature $T_C$. \cite{hwang,garcia}
Finally, due to the complex interplay of electron, orbital, spin
and lattice degrees of freedom, strong tendencies toward phase
separation are present in these materials. \cite{dagottonew}

In recent papers, \cite{cata,perroni,perroni1} some of us have
shown on the basis of a variational approach that the interplay of
the electron-phonon ($el-ph$) interaction and the double- and
super-exchange magnetic effects can be important to explain the
experimentally observed tendency of manganite bulk to form
inhomogeneous magnetic structures near the phase boundaries.
Employing this scheme, spectral and optical properties in absence
and in presence of magnetic field have been derived in the regime
of $CMR$ finding good agreement with experimental data.

The situation for manganite films is more complicated. Maximum
$MR$ values in films are usually larger and at lower temperatures
than in equivalent bulk materials. \cite{jinnew,aarts} This
enhanced change of the resistivity of the films has immediately
suggested device applications based on the sensitivity to magnetic
fields. \cite{prellier} An essential issue for manganite films is
to understand the role of the strain due to lattice mismatch
between the substrate and the film. Indeed it has been found that
properties such as magnetoresistance, magnitude of the temperature
$T_C$, resistivity, magnetization,
\cite{jinnew,prellier1,koo,kwon,walter,razavi,rao1,shreekala}
transport and magnetic anisotropies, \cite{prellier2,klein} and
spin and orbital order structure \cite{izumi} are sensitive to the
epitaxial strain. These properties are different from the changes
induced by hydrostatic or chemical pressure, since in-plane strain
generally leads to an out-of-plane strain of different sign.
Moreover the effects induced by the substrate are able to
influence the tendency toward phase separation,
\cite{hong,babushkina,biswas,liam,becker} induce inhomogeneities
\cite{bibes} in films, and cause new electronic behavior not found
in bulk materials of the same composition. \cite{biswas1}
Actually, the strain affects so many quantities that it could be
used to control the properties of interest by depositing films on
various substrates, changing the deposition conditions and the
post-annealing procedure, and varying the thickness.
\cite{prellier,kwon1}

In this paper we extend the variational approach of the bulk in
order to deal with manganite films. The size of the film is taken
into account considering the system made of a finite number of
planes and imposing open boundary conditions along the
out-of-plane direction of growth. The thickness dependence has
been studied finding similar results in the weak $el-ph$ coupling
regime appropriate for large bandwidth systems such as
$La_{1-x}Sr_xMnO_3$ ($LSMO$) films and in the intermediate regime
suitable for $La_{1-x}Ca_xMnO_3$ ($LCMO$) films: reduction of the
critical temperature, decrease of the magnetization and increase
of the resistivity especially at low temperatures. For thin films
characterized by a thickness larger than the possible dead layer,
the calculated results are in good agreement with experimental
behaviors.

The strain in the film is simulated through modifications of
in-plane and out-of-plane hoppings, that are related to lattice
parameters measured in these systems. It is found that the
compressive strain leads to an enhancement of the critical
temperature, while tensile strain weakens the ferromagnetic phase.
When the dependence on the bond angle $\theta$ is properly taken
into account, it is shown that also for compressive strain a
reduction of the transition temperature can occur. The strain
introduces also anisotropies in the transport properties showing
the consistency of the theory with experimental data. Finally
substrate-induced variations of the $el-ph$ coupling are
considered pointing out that, together with finite size and strain
effects, they can strongly influence the subtle balance of
coexisting phases favoring insulating phases when the thickness of
the film decreases.

In section I experimental and theoretical issues about strain in
manganite films are discussed; in section II the variational
approach and the transport properties of the bulk are generalized
to deal with manganite films; in section III the thickness
dependence is analyzed; in section IV the strain effects on the
phase diagram and transport properties are calculated; in section
V changes of the $el-ph$ coupling induced by the substrate are
considered.

\section{Experimental and theoretical background}
In this section we discuss experimental and theoretical issues
about strain in manganite films.

In most cases tensile strain suppresses ferromagnetism and,
consequently, critical temperature $T_C$. On the contrary,
compressive strain should reduce the resistivity and shifts $T_C$
toward higher temperatures with respect to the transition
temperature of the bulk.   The observed strain effect is usually
interpreted within the double-exchange model, since the hopping
matrix element $t$ can be altered by epitaxial strain through the
change of the $Mn-O$ bond length $d$ and the bond angle $\theta$.
However, recent studies show that compressive strain does not
always lead to enhancement of $T_C$, \cite{koo,rao1} and some
anomalous results have been also reported for tensile strain.
\cite{gong,zhang1}

In order to explain unusual results in manganite films, an extra
mechanism mediated by the orbital state has been proposed.
\cite{zhang1} Furthermore the theoretical work concerning
manganite films has distinguished between uniform bulk strain and
biaxial strain effects on the Curie point. \cite{millisf} The
biaxial strain increases the Jahn-Teller splitting and favors
electron localization in $e_g$ levels, causing $T_C$ to decrease.
The agreement with experimental data is in sign and order of
magnitude pointing out that a Jahn-Teller coupling is a crucial
variable. \cite{millisf,chen,worledge} However, this model does
not account for the observed decrease of $T_C$ upon thickness
reduction mainly in films of $LCMO$ on $SrTiO_3$ ($STO$) and
$LaAlO_3$ ($LAO$). \cite{rao1,bibes}

When the properties of epitaxial films are studied as a function
of thickness, different and complex features are usually observed.
First, a reduction of the Curie point $T_C$ and the $MI$
transition temperature $T_P$ of $LSMO$ and $LCMO$ films has been
reported when thickness decreases. This occurs both for films with
a gradually relaxed structure \cite{walter,rao1,wang} and for
fully strained films \cite{bibes} showing that the strain cannot
be the only factor responsible for the reduction of $T_C$ in very
thin films. Second, when thickness is reduced, the resistivity
increases and the magnetic moment often decreases. This has been
usually interpreted as due to the presence of a dead layer located
at interfaces. \cite{walter,bibes,sun,borges} The thickness of
these dead layers is of the order of a few $nm$ depending on the
substrate. The decrease of the Curie temperature and the increase
of resistivity are gradual suggesting that regions with higher
resistivity could be present at distances from the interface
larger than size of the dead layer. Third, changes in the phase
coexistence linked to the thickness dependence can lead to charge
trapping in thin films. \cite{bibes} Thus the thickness dependence
and the strain effect are far from being fully understood and
challenging.

\section{Variational approach and transport properties}

The adopted model takes into account the double-exchange
mechanism, the coupling of the $e_g$ electrons to lattice
distortions and the super-exchange interaction between neighboring
localized $t_{2g}$ electrons. \cite{cata,perroni,perroni1} Within
the single orbital approximation (reasonable in the doping regime
where $CMR$ occurs), the Hamiltonian reads

\begin{eqnarray}
H=&&\sum_{i,\delta} \left(\frac{S^{i,i+\delta}_0+1/2}{2
S+1}\right) t_{\delta} c^{\dagger}_{i} c_{i+\delta}
 +\omega_0 \sum_{i}a^{\dagger}_{i}a_{i}
+g \omega_0 \sum_{i} c^{\dagger}_{i}c_{i} \left(
a_{i}+a^{\dagger}_{i} \right)
  \nonumber \\
&& + \epsilon \sum_{<i,j>} \vec{S}_{i} \cdot \vec{S}_{j} - \mu
\sum_{i} c^{\dagger}_{i} c_{i}. \label{1r}
\end{eqnarray}
The quantities in eq.(\ref{1r}) have been discussed in previous
papers. \cite{cata,perroni} In films the electron transfer element
is assumed to be depending on the direction of hopping and open
boundary conditions are imposed along the $z$ axis assumed as the
out-of-plane axis of growth. As in the bulk, the dimensionless
parameter $\lambda=\frac{g^2 \omega_0 }{6 t} $ is introduced to
measure the strength of the $el-ph$ interaction in the adiabatic
regime.

Two canonical transformations are performed in order to treat the
$el-ph$ interaction variationally. Then the Bogoliubov inequality
is employed in order to derive the variational free energy of the
system using a test Hamiltonian characterized by free electron,
phonon and spin degrees of freedom. \cite{perroni} The electron
free energy per site reads

\begin{equation}
f_{test}^{el} = \frac{(-T)}{(N_x N_y N_z)}  \sum_{\bf{k}}
\log\left[1+e^{-\beta \xi_{\bf{k}}} \right]+\mu \rho,
\label{fro}
\end{equation}
where $T$ is the temperature, $\beta$ the inverse of $T$, $N_i$
the number of sites along the $i$ axis, and $\rho$ the electron
density. In eq.(\ref{fro}) we have $\xi_{\bf{k}} =
\bar{\varepsilon}_{\bf{k}} -\mu $, where $
\bar{\varepsilon}_{\bf{k}} = \varepsilon_{\bf{k}}+\eta $ is the
renormalized electronic band. The band dispersion
$\varepsilon_{\bf{k}}$ is
\begin{equation}
\varepsilon_{\bf{k}}=\varepsilon_{k_x,k_y}+\varepsilon_{k_z}=-2
t_{eff}^x [cos(k_x)+ cos(k_y)] -2 t_{eff}^z cos(k_z), \label{11ur}
\end{equation}
where the effective transfer integrals $t_{eff}^x$ and $t_{eff}^z$
take into account the polaronic and double-exchange effects, and
the quantity $\eta$ measures the electronic band shift due to the
$el-ph$ interaction. Due to the finite size along the $z$ axis,
the values of $k_z$ are given by
\begin{equation}
k_z= \frac{m \pi}{(N_z+1)},
\end{equation}
with $m=1,..,N_z$. The thermodynamic limit is performed along $x$
and $y$ directions, using a constant electronic density of states
$g_{2D}(\varepsilon)=1/(8 t_{eff}^x )$, that represents a simple
approximate expression for the exact density of states used for a
$2D$ lattice.

Within the variational approach, in a way analogous to the bulk
case, the linear response to an external electromagnetic field of
frequency $\omega$ and, consequently, the transport properties in
the limit $\omega \rightarrow 0$ can be calculated.
\cite{perroni,perroni1} The focus is on the real part of the
conductivity tensor $\Re \sigma_{\alpha,\gamma}(\omega)$ given by

\begin{equation}
\Re \sigma_{\alpha,\gamma}(\omega)= - \frac{ \Im
\Pi_{\alpha,\gamma}^{ret}(\omega) }{\omega}, \label{54r}
\end{equation}
where the $\Pi_{\alpha,\gamma}^{ret}$ is the retarded
current-current correlation function. It can be deduced making the
analytic continuation of the correlation function defined in
Matsubara frequencies as
\begin{equation}
\Pi_{\alpha,\gamma}(i \omega_n)=  - \frac{1}{(N_x N_y N_z)} \int_0
^{\beta} d\tau e^{i \omega_n \tau} \langle T_{\tau}
j^{\dagger}_{\alpha}(\tau) j_{\gamma}(0) \rangle,
 \label{55r}
\end{equation}
where the current operator $j_{\alpha}$ suitable for the film
geometry is

\begin{equation}
j_{\alpha}=i e \sum_{i,\delta} \delta t_{\alpha} \left(
\frac{S_0^{i+\delta \hat{\alpha},i}+1/2}{2 S+1} \right)
c^{\dagger}_{i+\delta \hat{\alpha}} c_i,
 \label{56r}
\end{equation}
with $e$ electron charge.

We perform the two canonical transformations and make the
decoupling of the correlation function in electron, phonon and
spin terms through the introduction of the test mean-field
hamiltonian.  In manganite films the electron correlation function
is

\begin{eqnarray}
&& \langle T_{\tau} c_i^{\dagger}(\tau) c_{i+\delta \hat{\alpha}}
(\tau) c_{i^{\prime}+\delta^{\prime} \hat{\gamma}}^{\dagger}
c_{i^{\prime}} \rangle_t=
\nonumber \\
- \sum_{{\bf k},{\bf k}_1} \Phi^{*}_{{\bf R}_i}({\bf k}) &&
\Phi_{{\bf R}_i + \delta \hat{\alpha}}({\bf k}_1) \Phi^{*}_{ {\bf
R}_{i^{\prime}}  + \delta^{\prime} \hat{\gamma}}({\bf k}_1)
\Phi_{{\bf R}_{i^{\prime}}} ({\bf k}) { \mathcal G}^{(0)} ( {\bf
k},-\tau ) { \mathcal G}^{(0)} ( {\bf k}_1,\tau ),
 \label{62r}
\end{eqnarray}
where the function $\Phi_{{\bf R}_i}({\bf k})$ is
\begin{equation}
\Phi_{{\bf R}_i}({\bf k})= \left( \frac{e^{i k_x i_x}}{\sqrt{N_x}}
\right) \left( \frac{e^{i k_y i_y}}{\sqrt{N_y}} \right)
\phi_{iz}(k_z),
\end{equation}
with $\phi_{iz}(k_z)$ given by
\begin{equation}
\phi_{iz}(k_z)=\frac{2 sin(k_z i_z)
}{\sqrt{(2N_z+1)-sin[(2N_z+1)k_z]/sin(k_z) }},
\end{equation}
and ${ \mathcal G}^{(0)} ( {\bf k},\tau )$ free electron Green's
function. To derive the optical properties, second order
fluctuations on the mean-field approach are fundamental since they
introduce scattering between charge carriers. \cite{perroni} The
effect of the damping can enter the calculation substituting $
{\cal G}^{(0)}$ for $ \tilde{{\cal G}}$ that in Matsubara
frequencies is expressed as
\begin{equation}
\tilde{ {\mathcal G} } ( {\bf k},i \omega_n )= \int_{- \infty}^{+
\infty} \frac{ d \omega } {2 \pi} \frac { \tilde{ A }( {\bf k},
\omega )} {i \omega_n - \omega } \label{38r},
\end{equation}
where the spectral function $\tilde{ A }$ is assumed to be

\begin{equation}
\tilde{ A }( {\bf k}, \omega )= \frac {\Gamma( {\bf k})} {
[\Gamma( {\bf k})]^2/4 +(\omega - \xi_{{\bf k}} )^2   },
\label{43r}
\end{equation}
with $\Gamma( {\bf k})$ rate of the scattering with optical
phonons and spin fluctuations evaluated on the energy shell.
\cite{perroni}

The conductivity tensor $\Re \sigma_{\alpha,\gamma}(\omega)$  is
deduced following the lines of our previous works.
\cite{perroni1,perroni2,perroni3} In the film geometry this tensor
is diagonal with $\Re \sigma_{x,x}(\omega)=\Re
\sigma_{y,y}(\omega)$, thus, in the limit $\omega \rightarrow 0$,
two different resistivities are obtained: in-plane resistivity
$\rho_{x,x}$ and out-of-plane resistivity $\rho_{z,z}$.

\section{Thickness dependence}

In this section we analyze the effects of the thickness on the
phase diagram, the magnetization and the transport properties of
the film. For the moment we do not introduce hopping anisotropies
being $t=t^i$, with $i=x,y,z$.

First the role of the size is investigated in the regime of weak
$el-ph$ coupling appropriate for $LSMO$ films. In Fig. 1 the phase
diagram of the film is reported for the bulk (solid line), for a
film made of 10 planes (dashed line) and 5 planes (dotted line).
For these values of the parameters the system exhibits a
continuous transition from a double-exchange ferromagnet to a
metallic paramagnet. A decrease of size leads to a reduction of
the ferromagnetic phase that seems to be a generic feature of
magnetic films. \cite{zhang2} Actually the critical temperature
shifts to lower temperatures than that of the bulk when the
spin-spin correlation exceeds the film thickness. \cite{fisher} In
the inset of Fig.1 the variation of the Curie temperature as
function of the number of the planes is reported. The critical
temperature shows a large decrease as the number of planes is
reduced. This effect is very pronounced when the number of planes
is smaller than $10$: in fact in this case there is a shift of the
critical temperature larger than $10 \%$ of the corresponding
temperature of the bulk.

We have investigated the thickness dependence also for a manganite
film in the regime of intermediate $el-ph$ coupling and bandwidth
suitable for $LCMO$ films. In this case near the $MI$ transition
the system segregates in ferromagnetic and antiferromagnetic or
paramagnetic domains of itinerant and localized carriers,
respectively. \cite{cata,perroni,perroni1} In Fig. 2 the phase
diagram for these systems is shown in correspondence with
different sizes of the system. We find that the reduction of the
ferromagnetism does not depend on the order of the transition.
Although the transition lines are different from the phase diagram
of the film in the weak-coupling regime, the decrease of the
ferromagnetic phase is of the same order. This is confirmed also
by the behavior of the critical temperature as a function of the
number of planes, as reported in the inset of Fig. 2. Even if the
transition temperatures are lower, the shape of the curve bears a
strong similarity with the corresponding plot for systems in
weak-coupling regime: in both cases the shift of Curie temperature
$S(N)=1-T_C(N)/T_C(\infty)$ follows a power law $S(N)\sim N^{-r}
$. \cite{zhang2}

We note that, in manganite films grown without detectable lattice
defects, the decrease of the critical temperature occurs for a
thickness smaller than $20 nm$. \cite{walter} Therefore for thin
films characterized by thickness larger than the possible dead
layer, the size effect can be relevant in the interpretation of
experimental behaviors allowing to explain the gradual decrease of
the critical temperature with decreasing film size. Actually, the
tendency towards the localization in the thinnest films cannot be
entirely ascribed to the presence of a dead layer.

Due to the thickness effect, the ground state presents a weaker
order of spins, that can be easier destroyed with increasing
temperature. Indeed, as the size of the film is reduced, the
magnetization assumes smaller values at low temperatures.
\cite{borges} In a double-exchange model a weakening of the
ferromagnetism is expected to reduce the mobility of the carriers.
This is confirmed by the calculated in-plane resistivity
$\rho_{x,x}$ that, as reported in Fig. 3, at $T=0.05 \omega_0$,
exhibits a significant increase as the number of planes decreases
(the out-of-plane resistivity $\rho_{z,z}$ shows a similar
behavior). Like the critical temperature, large variations occur
when the film is very thin suggesting that size effect can be
important to explain the gradual increase of the resistivity as
the thickness decreases. However, the behavior of resistivity with
respect to the number of planes is different as the temperature
increases. In fact, as shown in the inset of Fig. 4, at higher
temperatures the resistivity of the film is smaller for thinner
films. This can be understood considering that the differences in
the spin order due to size effects diminish with increasing
temperature. So the bare behavior of the scattering rate with
optical phonons emerges: this type of scattering decreases when
the size of the system is reduced in agreement with what happens
in quantum wells. \cite{mitin} At high temperatures, in addition
to size effect, other effects such as the strain have to be
considered in order to get a better agreement with experimental
data.

\section{Strain effects}

In this section we deal with strain effects on the properties of
the film.

The strain in the film is introduced through the modifications of
in-plane and out-of-plane hoppings, that are linked to lattice
parameters measured in these systems. It can be assumed $t^x=t^y
\propto cos(\phi_{in})/d_{in}^{3.5}$, where $d_{in}$ is the
in-plane bond length and $\phi_{in}=(\pi-\theta_{in})/2$, with
$\theta_{in}$ in-plane bond angle. \cite{chen,medarde} In an
analogous manner, we can have $t^z \propto
cos(\phi_{out})/d_{out}^{3.5}$, where $d_{out}$ is the
out-of-plane bond length and $\phi_{out}=(\pi-\theta_{out})/2$,
with $\theta_{out}$ out-of-plane bond angle. If one neglects the
angular dependence, the ratio of the hopping amplitudes of the
film with respect to the bulk values are given by
$t^x/t_B=(a_B/a)^{3.5}$ and $t^z/t_B=(c_B/c)^{3.5}$, where
$a_B=c_B$ is the lattice parameter of bulk, $a$ and $c$ are
in-plane and out-of-plane lattice parameters of the film measured
by X-ray diffraction experiments.
\cite{koo,rao1,suzuki,ju,miniotas} These ratios simulate the
strain of the film within the variational approach, so its effects
on the phase diagram and the transport properties can be
estimated.

The variations of the phase diagram are controlled mainly by the
changes of in-plane hopping, so that compressive strain ($c/a
>1$) leads to an enhancement of the critical temperature, while
tensile strain ($c/a<1$) to a reduction. The increase of the
compressive (tensile) strain is able to perturb the system also at
low temperatures where the ferromagnetic alignment of the spins is
reinforced (reduced). Deriving in-plane and out-of-plane hoppings
from the lattice parameters of a compressively strained $LSMO$
film,\cite{suzuki} it is found that the strain introduces
anisotropies in the transport properties of the films in the weak
coupling regime. \cite{klein} Indeed, as shown in Fig. 4, the
in-plane resistivity $\rho_{x,x}$ decreases in temperature with
increasing compressive strain, while the out-of-plane resistivity
$\rho_{z,z}$ increases. This is explained by the fact that,
increasing the strain, the in-plane hopping is enhanced, while the
out-of-plane hopping is reduced. In any case, at low temperatures,
due to the enhancement of the ferromagnetic order induced by the
compressive strain, both the resistivities decrease with
increasing strain.

Next we have investigated the effects of compressive and tensile
strain on the phase diagram of a manganite film in the
intermediate coupling regime. The in-plane and out-of-plane
hopping amplitudes have been deduced considering the lattice
parameters of $LCMO$ films grown on different substrates.
\cite{koo} The film on $LAO$ corresponds to the ratio $c/a=1.009$,
that on $MgO$ to $c/a=1.002$, finally that grown on $STO$ to
$c/a=0.981$. As shown in Fig. 5, in the first case, the
compressive strain (dotted line with diamonds) leads to an
enhancement of the critical temperature with respect to the bulk
value (solid line with circles) unlike experimental results.
\cite{koo} In the second case, even if field strain is near the
unity (dash-dotted line with down triangles), the critical
temperature shifts at lower temperatures in agreement with
experimental data. Actually in this case both the lattice
parameters exceed the bulk value, determining a reduction of the
in-plane hopping. In the third case, for tensile strain (dashed
line with squares), the shift of the Curie temperature in
comparison with the bulk value is consistent with experimental
measurements. \cite{koo}

We stress that the results discussed above have been obtained
neglecting the dependence of hopping amplitudes on the $Mn-O-Mn$
bond angle $\theta$. However, in bulks of $La_{1-x}A_xMnO_3$
perovskites, it has been shown that changes in $A$ site ionic size
affect only $\theta$, while the bond length $d$ seems to be
unchanged ($d^*=0.196 nm $) suggesting as a maximum lattice
parameter $2 d^*=0.392 nm$. \cite{garcia} Recent measurements in
$LCMO$ thin films on $LAO$ and $STO$ substrates \cite{miniotas}
have showed that $d_{in}$ is fixed to a value that does not depend
on the type of strain, while the in-plane bond angle is changed.
The angle $\theta_{in}$ becomes larger (smaller) under tensile
(compressive) strain, showing consistency with the elongation
(contraction) of $a$. Similar results have been observed also in
$LSMO$ films on $LAO$. \cite{ju} Hence, it is reasonable to assume
that values of the lattice parameter larger than those of the bulk
imply stretching of $d$ ($\theta$ near $\pi$) and smaller lattice
parameters induce a contraction of $\theta$ ($d$ near $d^*$).
Thus, for $LCMO$ films grown on $LAO$ the angular dependence is
not negligible. \cite{yuan} Actually, deriving $\theta_{in}$ from
experimental data \cite{miniotas} ($\theta_{in} \simeq
152.5^{\circ}$) and considering only the changes of in-plane
parameters (that are the most important), we can deduce the
variation of the in-plane hopping from the knowledge of the
$Mn-O-Mn$ angle of the bulk \cite{dai} ($\theta \simeq
160^{\circ}$). As reported in Fig. 5 (dash-dotted line with up
triangles), even if the strain is compressive, the critical
temperature is reduced compared with bulk value in agreement with
experimental data. \cite{koo}

\section{Variations of the $el-ph$ coupling}
In this section we focus our attention on the variations of the
$el-ph$ coupling linked to the presence of strain that can be
included in order to improve the description of manganite films.

Both uniform bulk strain and biaxial strain can affect the $el-ph$
coupling strength of the carriers in manganite films.
\cite{millisf} While the biaxial distortions tend to increase the
localization of the carriers, the bulk strain can lead to an
increase or a decrease of $el-ph$ coupling depending on the sign
of the strain. Actually, a uniform compression tends to increase
the electron hopping amplitudes reducing the importance of the
$el-ph$ coupling. So, unlike tensile strain, for compressive
strain two opposite behaviors influence the variations of the
$el-ph$ coupling. In any case, for very thin films, the tendency
towards localization can be simulated in our scheme by an
enhancement of the $el-ph$ coupling.

For $LSMO$ films grown on $STO$ the size and strain dependence of
the transition temperature is strongly influenced by the
strain-induced $el-ph$ coupling. \cite{chen} Actually we have
verified that a small reduction (increase) in $el-ph$ coupling can
induce a large enhancement (decrease) of the critical temperature
in the $CMR$ regime.

For fully strained $LCMO$ films grown on different substrates the
delicate balance of segregating phases is strongly influenced by
finite size and strain effects. Deducing sizes and strain fields
of the $LCMO$ films grown on $STO$, \cite{bibes} in Fig. 6 we show
the corresponding phase diagrams calculated supposing a likely
distribution of the quantity $Rg$, that represents the ratio
between the $el-ph$ coupling $g$ of the film and that of the
thickest film. The concomitant effects induce a strong lowering of
the transition temperature and at low $T$ an increase in the
weight of antiferromagnetic insulating phases. In fact, for a film
characterized by a thickness of $6$ planes and a ratio $Rg=1.05$,
an antiferromagnetic insulating phase of localized charges is
stabilized at low temperatures in agreement with experiment.
\cite{bibes} Therefore, the insulating dead layer could originate
as natural consequence of reduced size of the film, strain effect
and increased $el-ph$ coupling. In Fig. 7 the magnetization (upper
panel) and the in-plane resistivity (lower panel) are reported as
function of the temperature for three different values of the
strain and $el-ph$ coupling. As result of the competition between
ferromagnetic metallic and antiferromagnetic or paramagnetic
insulating phases, \cite{liam,perroni,perroni1,perroni2} the
magnetization is strongly reduced, while the resistivity is
largely enhanced showing consistency with experiments
\cite{bibes}. Finally we stress that the nearly constant behavior
of the resistivity at intermediate temperatures is characteristic
of the antiferromagnetic phase of localized carriers. Only at
higher temperatures, in the paramagnetic phase, the resistivity
strongly diminishes with increasing temperature.

\section{Summary and conclusions}
In this paper we have analyzed the thickness dependence and strain
effects in films of $La_{1-x}A_xMnO_3$ perovskites within the
$CMR$ regime.

A variational approach previously proposed for manganite bulk has
been generalized in order to consider the film geometry. A
reduction in the thickness of the film causes a decrease of the
transition temperature and the magnetization, and increase of the
resistivity especially at low temperatures. If the film is very
thin but has thickness larger than the possible dead layer, the
calculated results show good agreement with experimental
behaviors. The strain is associated with the changes of in-plane
and out-of-plane electron hopping amplitudes induced by the
substrate. The variations of the phase diagram with respect to the
bulk are controlled mainly by the changes of in-plane hopping, so
that the compressive strain leads to an enhancement of the
critical temperature, while tensile strain weakens the
ferromagnetic phase. If the dependence of the hopping matrix
elements on the $Mn-O-Mn$ bond angle is properly considered, the
strain effects on the transition temperature and transport
properties are consistent with experimental data. Finally it is
shown that substrate-induced variations of the $el-ph$ coupling,
together with finite size and strain effects, can strongly
influence the subtle balance of coexisting phases favoring
insulating phases when the thickness of the film decreases.

Although the results presented in this paper are quite
satisfactory, we wish to mention other mechanisms that could
influence the thermodynamic and transport properties, at least
qualitatively. First a strain-induced modification of $e_g$
electron orbital stability can affect the phase diagram of the
system. \cite{zhang1,fang,konishis,shiraga,calderon} Changes in
oxygen content resulting in cationic vacancies due to the
annealing may shift the critical temperature at values much higher
than any bulk values in the series compounds.
\cite{prellier1,shreekala} The loss of magnetic moment and the
increase of resistivity could be caused by the domain-type
disorder. \cite{aarts,blamire} Even if in manganite films spatial
inhomogeneities can be intrinsic, \cite{becker} the disorder
induced by the substrate could be responsible also for phase
separation. \cite{dagottonew,bibes}

As stressed in the previous section, the tendencies toward phase
separation can be important to explain the presence of the
insulating dead layer in manganite films. The variational approach
used in this paper can be further generalized to consider
insulating planes separated through an interface from metallic
planes. Strain effects and thickness dependence could strongly
perturb the size of insulating and metallic planes giving rise to
the dead layer. Work in this direction is in progress.

\section*{Acknowledgments}
V. Cataudella, G. De Filippis, G. Iadonisi, C. A. Perroni, and F.
Ventriglia acknowledge financial support by PRIN "Electronic and
optical properties in manganite films".

\section*{Figure captions}
\begin{description}

\item  {F1}
Phase diagrams corresponding to $t=2.5  \omega_0$, $\lambda=0.48$
and $\epsilon=0.05 \omega_0$ for different numbers of film's
planes. $PM$ means Paramagnetic Metallic, $FM$ Ferromagnetic
Metal. In the inset the corresponding variation of the critical
temperature $T_C$ as a function of the number of the film's planes
at $x=0.3$. The temperatures are expressed in units of $\omega_0$.

\item  {F2}
Phase diagrams corresponding to $t=1.8 \omega_0$, $\lambda=0.65$
and $\epsilon=0.01 \omega_0$ for different numbers of film's
planes. $PI$ stands for Paramagnetic Insulating. In the inset the
resulting variation of the critical temperature $T_C$ as a
function of the number of the film's planes at $x=0.3$. The
temperatures are expressed in units of $\omega_0$.

\item  {F3}
The resistivity $\rho_{x,x}$ as a function of the number of planes
for $T=0.05 \omega_0$ at $x=0.3$ and $t=2.5 \omega_0$. In the
inset the variation in temperature (in units of $\omega_0$) of the
resistivity $\rho_{x,x}$ for different numbers of planes (the
arrows indicate the transition temperatures). The resistivity is
in units of $\frac{m \omega_0}{e^2 c}$, where $c$ is the hole
concentration and $m= \frac {1}{2t}$.

\item  {F4}
(a) The resistivity $\rho_{x,x}$ as a function of the temperature
(in units of $\omega_0$) for different values of strain at $t=2.5
\omega_0$ and $x=0.3$.

(b) The resistivity $\rho_{z,z}$ as a function of the temperature
(in units of $\omega_0$) for different values of strain at $t=2.5
\omega_0$ and $x=0.3$.

The resistivities are expressed in units of $\frac{m \omega_0}{e^2
c}$, where $c$ is the hole concentration and $m= \frac {1}{2t}$
and the arrows indicate the transition temperatures.

\item  {F5}
Phase diagrams corresponding to $t=1.8 \omega_0$, $\lambda=0.65$
and $\epsilon=0.01 \omega_0$ for different values of strain. $PI$
means Paramagnetic Insulating. The temperatures are expressed in
units of $\omega_0$.

\item  {F6}
Phase diagrams corresponding to $t=1.8 \omega_0$, $\lambda=0.68$
and $\epsilon=0.02 \omega_0$ for different values of size, strain
and $el-ph$ coupling. $PI$ means Paramagnetic Insulating. The
temperatures are expressed in units of $\omega_0$.

\item  {F7}
Magnetization (in units of the saturation magnetization) and
resistivity (in units of $\frac{m \omega_0}{e^2 c}$, wit $c$ hole
concentration and $m= \frac {1}{2t}$) as a function of the
temperature (in units of $\omega_0$) at $t=1.8 \omega_0$ and
$x=0.3$ for different values of size, strain and $el-ph$ coupling.

\end{description}

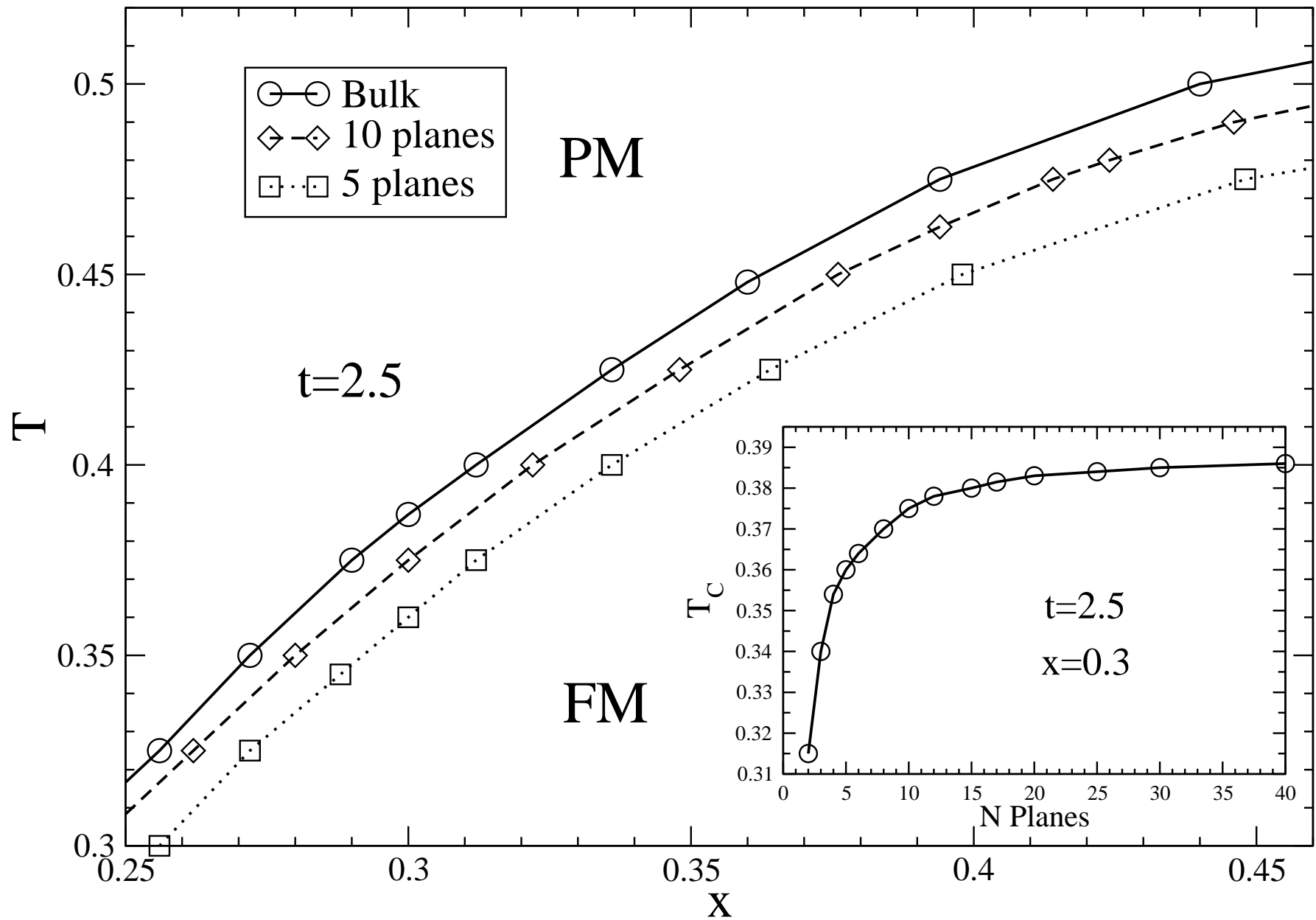
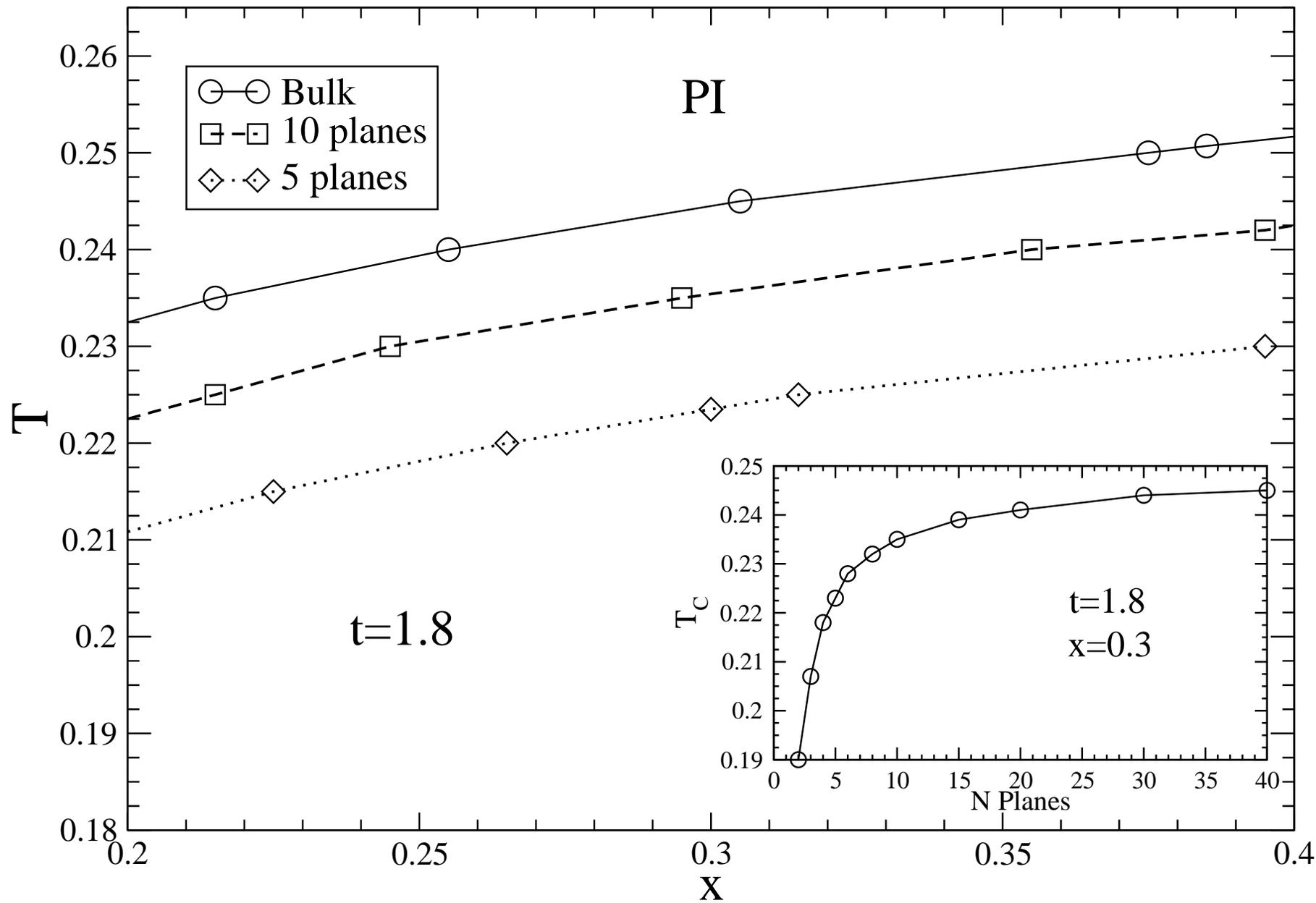
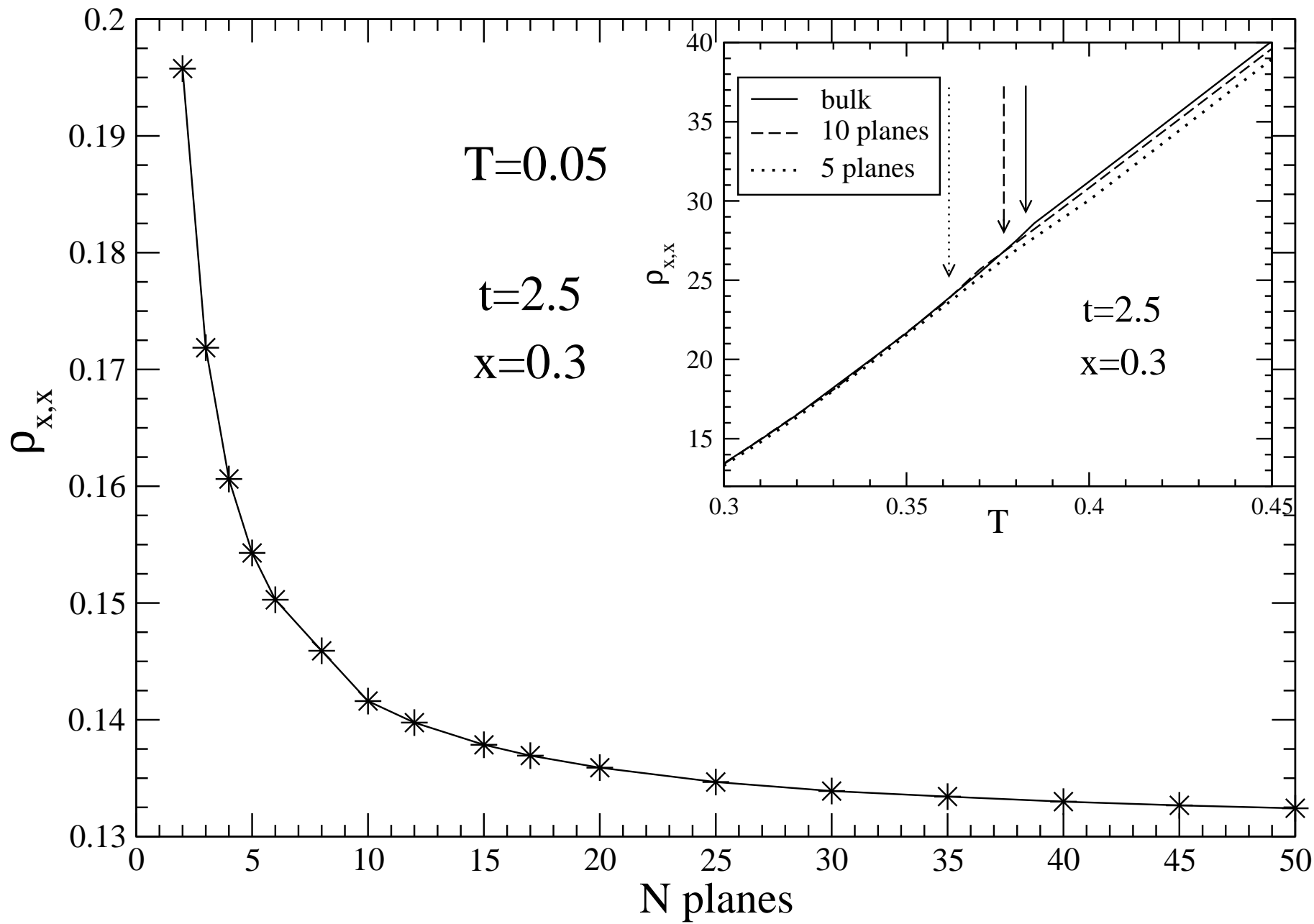
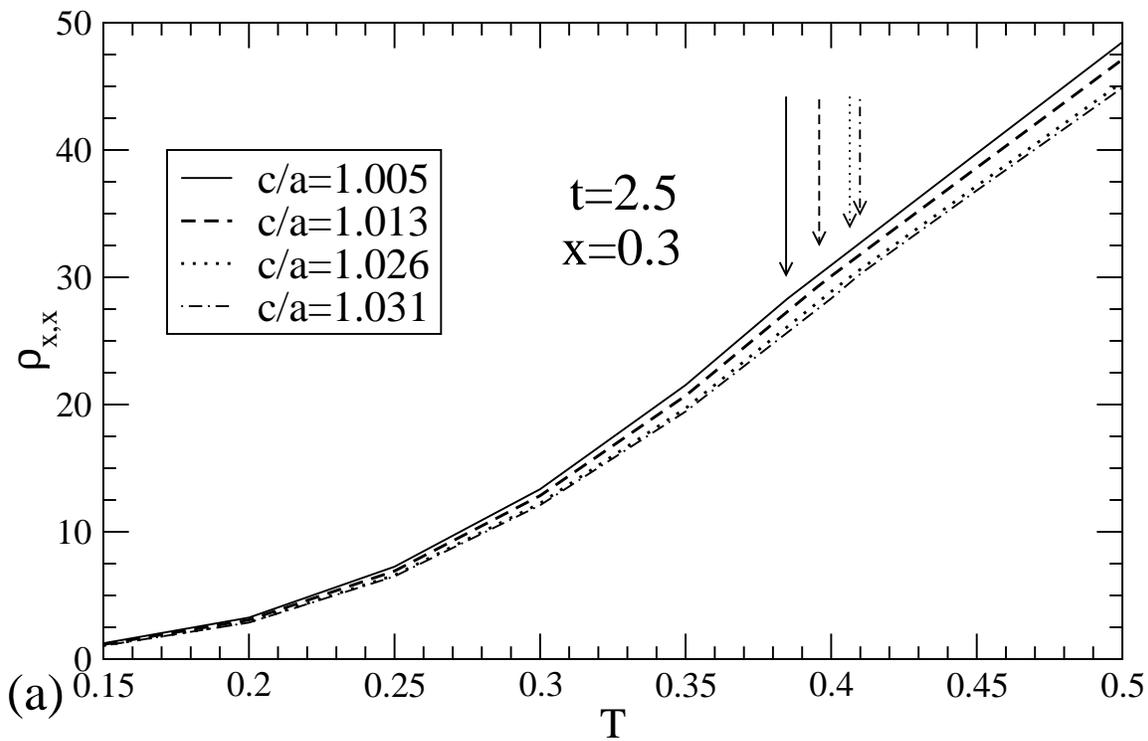
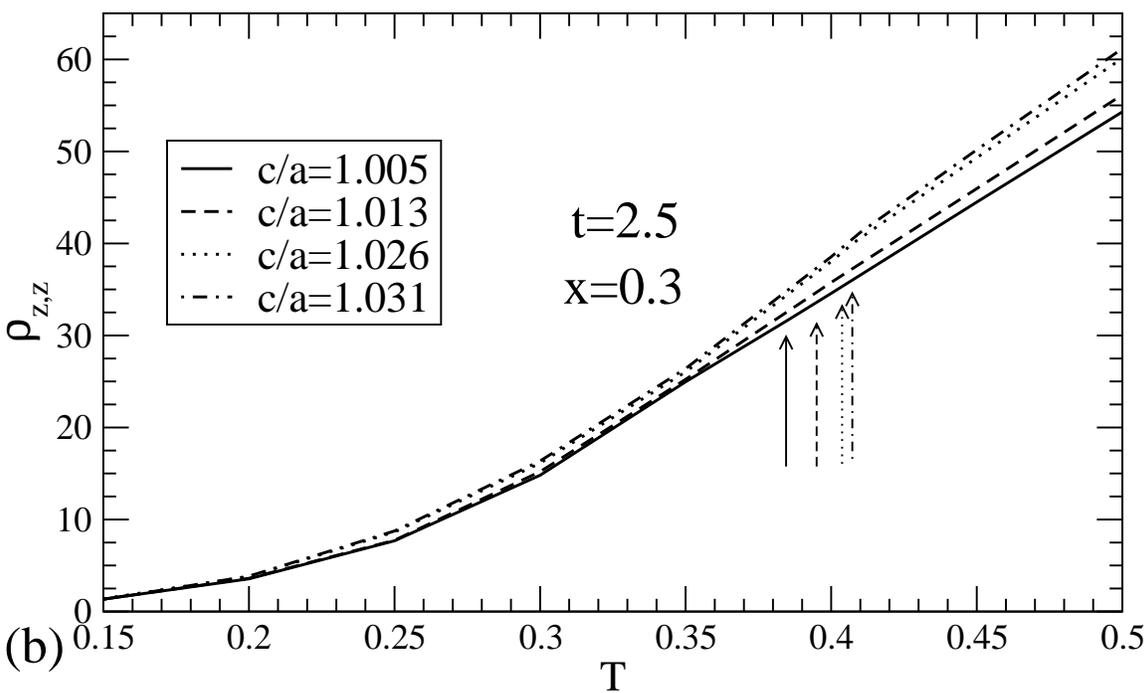
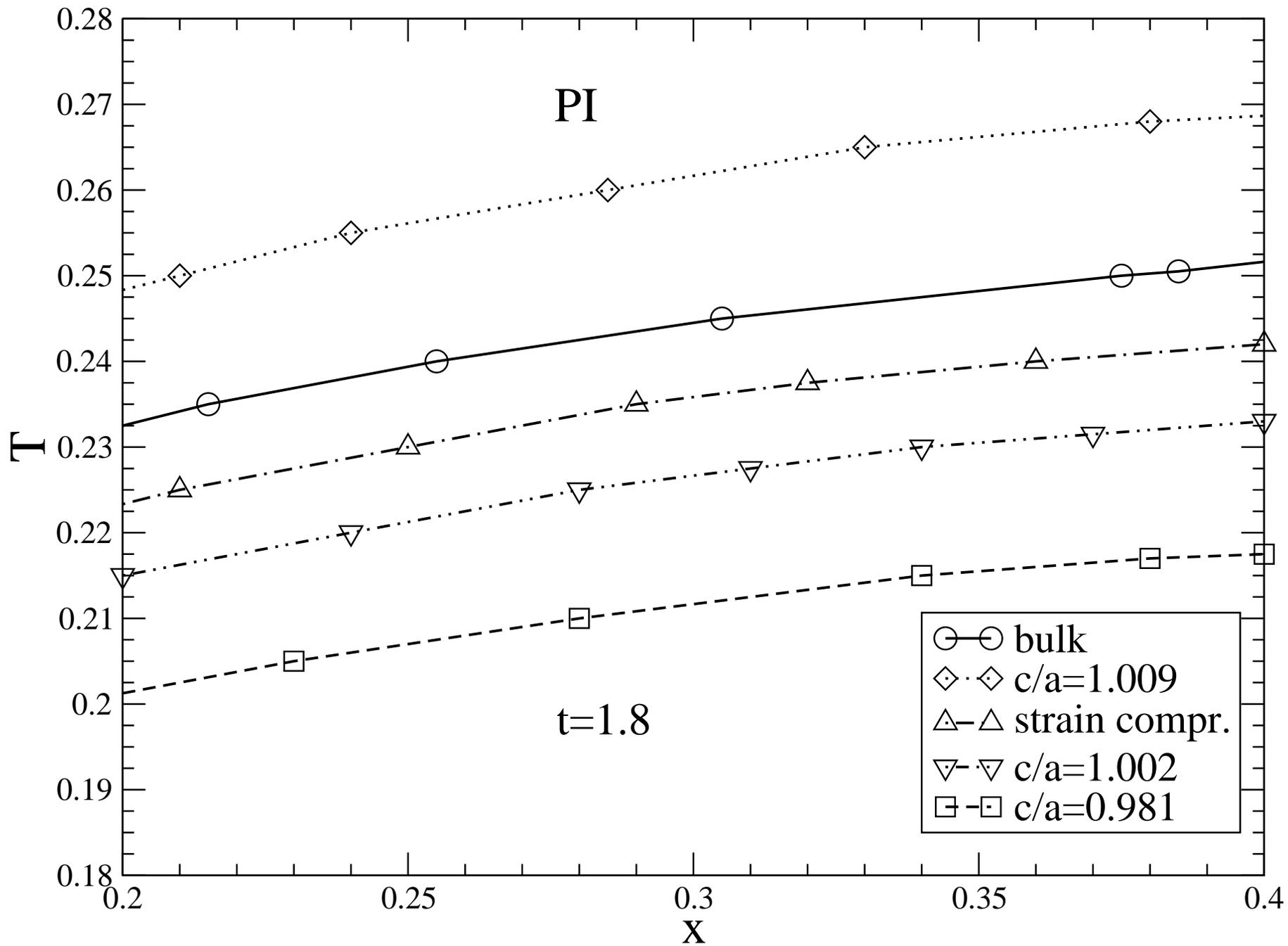
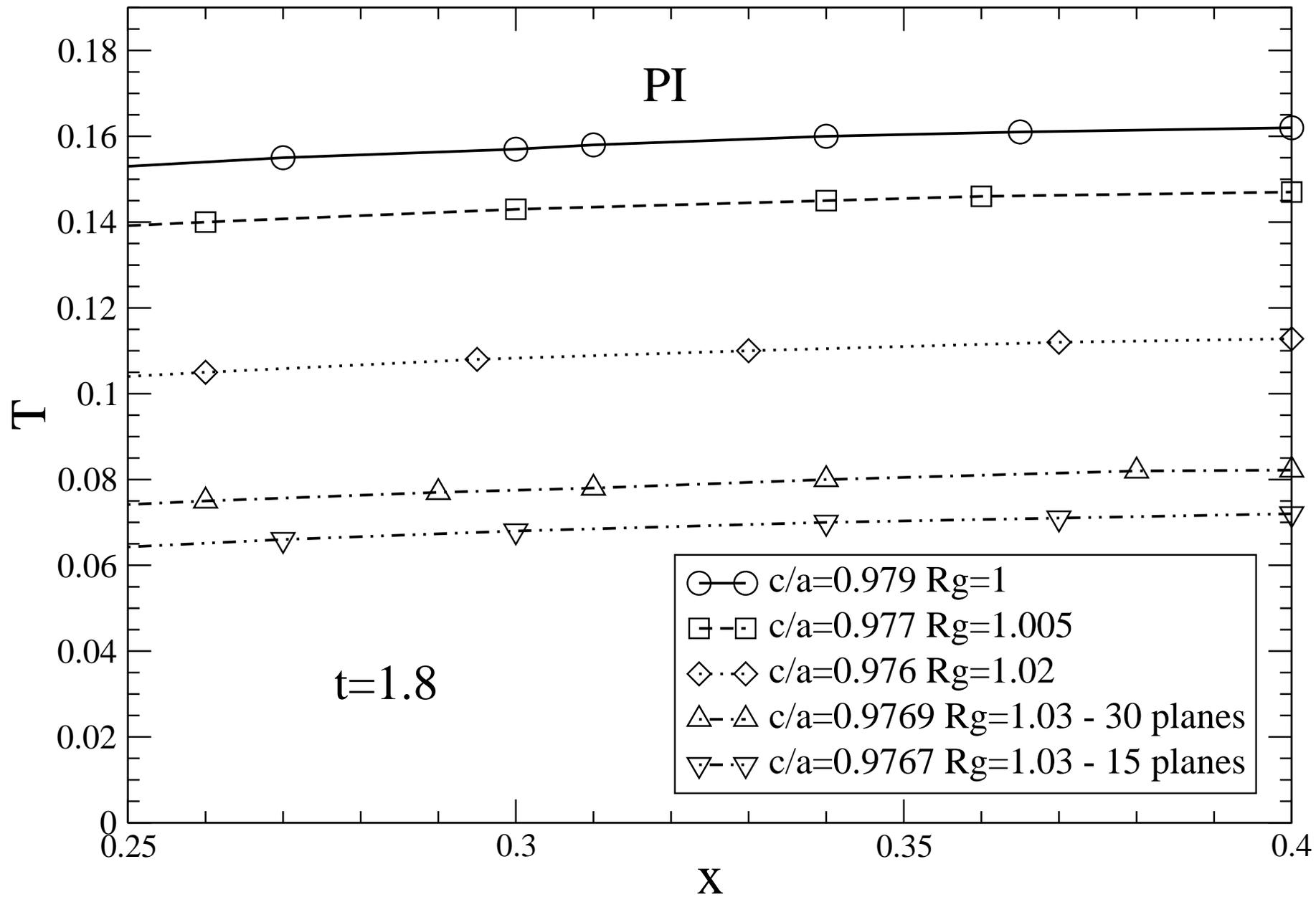
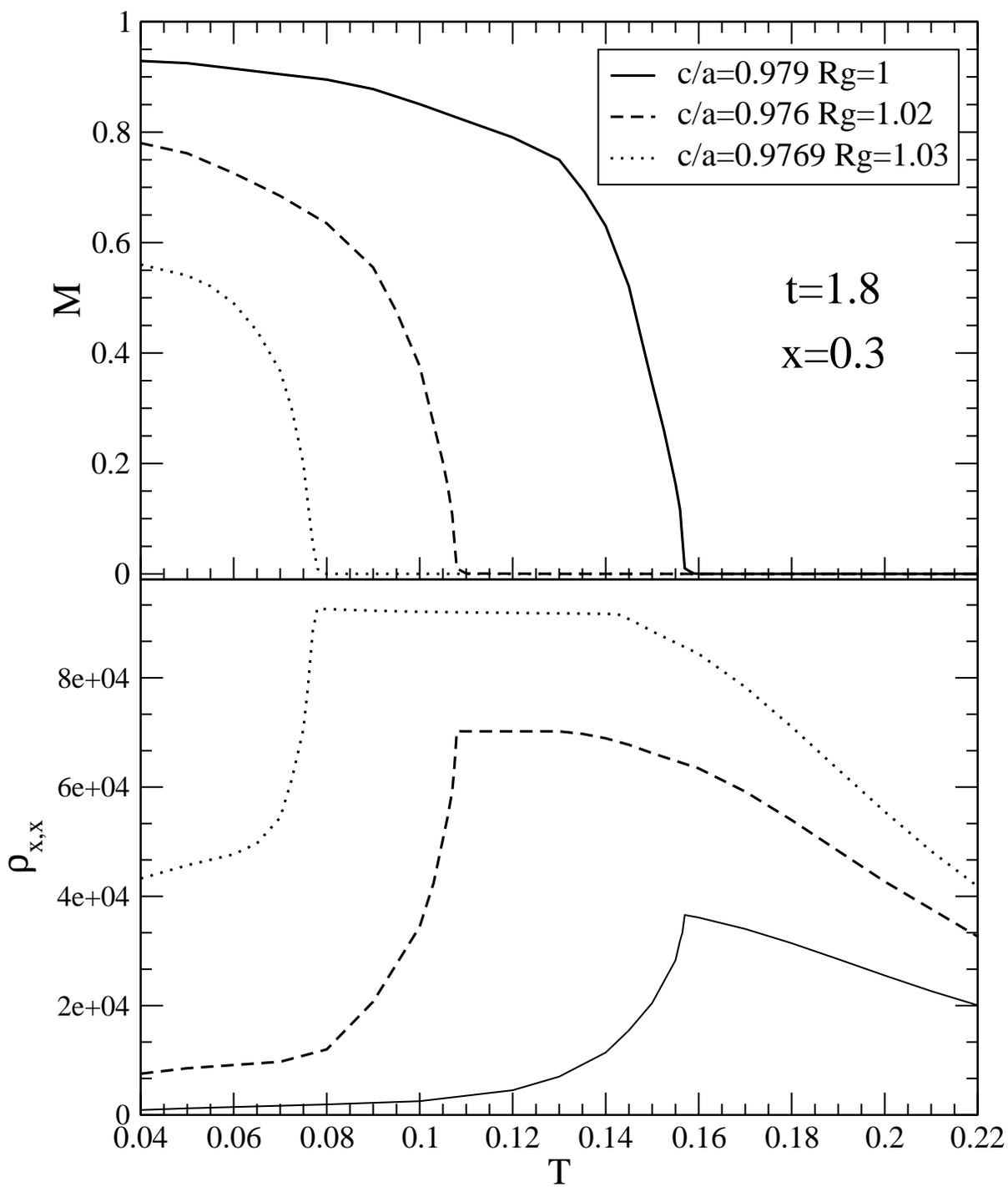
\end{document}